\documentclass[12pt]{iopart}
\usepackage{graphicx}

\usepackage{epsfig}
\usepackage{bm}

\newcommand{\be}{\begin{equation}}
\newcommand{\ee}{\end{equation}}
\newcommand{\bea}{\begin{eqnarray}}
\newcommand{\eea}{\end{eqnarray}}
\newcommand{\nn}{\nonumber}
\newcommand{\ket}[1]{ \mid#1\rangle}
\newcommand{\bra}[1]{\langle#1\mid}
\newcommand{\abs}[1]{\left|#1\right|}
\newcommand{\laj}{\lambda_j}

\newcommand{\roj}{\abs{\rho_j}^2}
\newcommand{\sij}{\abs{\sigma_j}^2}
\newcommand{\gaj}{\abs{\gamma_j}^2}

\begin{document}
\title{Entanglement generation and perfect state transfer in ferromagnetic qubit chains}
\author{Giulia Gualdi,  Irene Marzoli, and Paolo Tombesi}
\address{Dipartimento di Fisica, Universit\`a degli Studi di Camerino, I-62032 Camerino, Italy }
\date{\today}

\begin{abstract}
We propose to use ferromagnetic systems for entanglement generation and distribution  together with perfect state transfer between distant parties in a qubit chain. The scheme relies on an effective 2-qubit dynamics,  realized by leaving two empty sites in a uniformly filled  chain. This allows long-range interacting qubit chains to serve as quantum channels for both tasks with optimal performances. Remarkably, the entanglement between sender and receiver sites is independent of both the transmission distance and the system size. This property opens new perspectives for short and mid-range quantum communication with qubit chains.
\end{abstract}

\maketitle
\section{Introduction}
Quantum tasks such as quantum teleportation require the generation of entanglement as well as the transfer of a quantum state between distant parties \cite{Bennett93}.
Several theoretical works have shown that antiferromagnetic systems are suited for generating and distributing entanglement between distant parties \cite{Campos06,Campos07,Amico08}. This capacity  is related to the specific ground state properties of the different antiferromagnetic systems. So far, all proposals have concentrated on nearest-neighbour interacting spin systems, due to the difficulty in finding the ground state for an arbitrary long-range interacting system. Antiferromagnetic systems allow to generate entangled states, but the amount of entanglement decreases with the distance between the members of the entangled pair.

On the other side ferromagnetic systems may serve as quantum channels for short and mid-range quantum communication \cite{Bose03}.
These systems are traditionally regarded as being capable of transmitting a quantum state  \cite{Bose03,Christandl04,Albanese04},  but as not being able of generating any entanglement between two distant parties \cite{Amico08}. In general, entanglement generation and perfect state transfer have not been put into direct relation for ferromagnetic systems. Rather, different classes of physical  systems have been proposed for accomplishing the two goals separately. One exception is represented by the ferromagnetic branched chain analyzed in Ref.~\cite{Amico07}. Here, one can generate
entanglement between two receivers by means of a state transfer protocol.  This implies that entanglement generation is not distinguishable from state transfer, so that the two tasks are not separately addressed.
Moreover the procedure is not scalable, since it relies on specifically engineered  qubit couplings \cite{Christandl04,Albanese04}.

In this paper we prove that starting from a $N$-qubit
 system,  whenever it is possible to restrict the dynamics to an effective
  2-qubit subspace, one obtains a quantum channel suitable
 for both
optimal state transfer and entanglement generation and distribution.
Remarkably enough our procedure applies to ferromagnetic systems as well.
According to  the
time at which the final measurement takes place one can
select, on the same quantum channel, one of the two
tasks -- state transfer or entanglement generation. Moreover, the generated entanglement  is independent of both system size  and transmission distance.

\begin{center}
\begin{figure}\begin{center}\begin{tabular}{c}
\includegraphics[width=8cm]{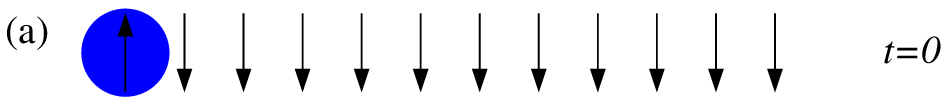}\\
\includegraphics[width=8cm]{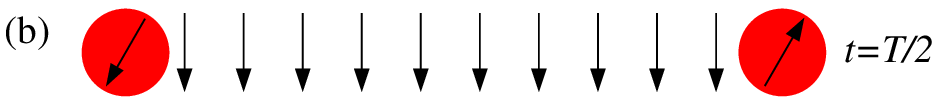}\\
\includegraphics[width=8cm]{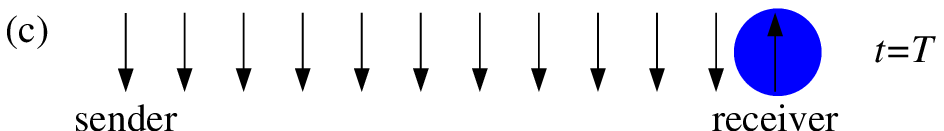}\end{tabular}\end{center}
\caption{\label{fig1} (Colour online) Time evolution of a ferromagnetic qubit chain:  the initial separable state with all spins down except the one at sender's site (a) evolves into a maximally entangled state between sender and receiver at $t=T/2$ (b), and finally  for $t=T$ is transferred to the other end of the chain (c). }
\end{figure}\end{center}
Without resorting neither to  dynamical control \cite{Haselgrove05}
nor to ancillary systems \cite{Burgarth05,Giovannetti06}
, perfect state transfer between two distant parties of  a ferromagnetic system is attained only if its spectrum is mirror-periodic \cite{Christandl04,Albanese04,Karbach05,Yung05,Kay06}. This means that the time-evolution maps an arbitrary quantum state into its mirror symmetric with respect to the centre of the system. For example, when the information is initially encoded at one end of  an open chain, after the so-called transfer time we will retrieve the message at the opposite end.
There are two possible strategies to achieve this goal.
The first one is to design  a $N$-qubit mirror-periodic system  by  carefully adjusting the coupling strengths between the members of a $N$-qubit array \cite{Christandl04,Albanese04,  Amico07,Karbach05,Yung05,Kay06}. The second one is to map a $N$-qubit system into an effective $2$-qubit system, which is by definition mirror-periodic. This task is accomplished by slightly detaching  sender and receiver from the channel \cite{Campos06, Gualdi08,Wojcik05, Plenio05}. ln this paper we show that a system, which exhibits perfect state transfer by means of this second approach, is also capable of generating maximally entangled states between distant parties. On the opposite, $N$-qubit mirror-periodic systems are prevented from reaching this goal.\\
This manuscript is organized as follows: in Sec. \ref{cond}
we derive analytically the conditions for generating maximally entangled states between the two ends of a generic $N$-qubit chain. We then discuss, in Sec. \ref{ent}, the relationship between entanglement generation and perfect state transfer, and suggest a scalable and straightforward-to-implement protocol to achieve both tasks. As an example, in Sec. \ref{heis}, we apply our strategy to a long-range interacting spin chain, described by the Heisenberg XYZ isotropic model. In Sec. \ref{num} we provide numerical results supporting our theoretical insight. Finally, we draw our conclusions in Sec. \ref{concl}.

\section{Conditions for the generation of maximally entangled states}\label{cond}

Consider a  $N$-spin system, whose Hamiltonian $H$ preserves the total magnetization $M=\sum_{i=1}^N\sigma_i^z$, where $\sigma^z_i$ is the $z$-component of the Pauli spin operator acting on the spin at the $i$th site. We consider the case of transmitting a single excitation from sender to receiver [Fig. \ref{fig1}(a)]. As demonstrated in Ref. \cite{Gualdi08},
perfect state transfer  for a generic system,  without a mirror-periodic spectrum, implies that  the  $N\times N$-dimensional Hilbert space $\mathcal{H}$ of the system
must be expressed as the sum of two disjoint subspaces, $\mathcal{H}^{2\times2}_{(s,r)}$ and $\mathcal{H}^{(N-2)\times(N-2)}_{channel}$, corresponding, respectively, to the sender-receiver subspace and to the rest of the chain, i.e. the channel.
 Ideally, perfect state transfer of an excitation from the sender site $s$ to the receiver site $r$ only involves a pair of eigenvectors
\be\ket{\lambda_{\pm}}=\frac{1}{\sqrt2}(\bm{\ket s}\pm\bm{\ket r}),\label{lapiu}
\ee where $\bm{\ket{i}}$ denotes the basis state with all the qubits in the $\ket{0}$ state  except the one at the $i$th site which is in $\ket 1$. In a truly 2-qubit space, restricted to sender and receiver only, these eigenvectors can be written as
 \be \ket{\lambda_{\pm}}=\frac{1}{\sqrt2}(\ket{10}\pm\ket{01})=\bm{\ket{\Psi^{\pm}}}.\label{bell}\ee
We note that the ideal system eigenvectors correspond to the two Bell states, $\bm{\ket{\Psi^{\pm}}}$,  with the same magnetization for both members of the entangled state.  Hence, perfect state transfer is achieved owing to the interplay between the two maximally entangled states $\bm{\ket{\Psi^{+}}}$ and $\bm{\ket{\Psi^{-}}}$. This observation suggests that  the same  system should be able to generate entanglement as well.

To this end, we now compute the concurrence \cite{Wootters98} as a measure of the entanglement between a generic sender-receiver pair of a $N$-qubit system. The system, initially prepared in the separable state \be\bm{\ket{\Psi(0)}}=\cos{\frac{\theta}{2}}\bm{\ket{0}}+e^{-i\phi}\sin{\frac{\theta}{2}}\bm{\ket{s}},\label{psio}\ee evolves according to the unitary time-evolution operator \be\bm{U}(t)=\exp(-iHt)\quad\mbox{with} \quad\hbar=1\ee into \be\bm{\ket{\Psi(t)}}=\bm{U}(t)\bm{\ket{{\Psi(0)}}}.\ee
Following Ref. \cite{Wootters98}, the concurrence is defined as $\mathcal{C}=\max\{0,\lambda_{1}-\lambda_{2}-\lambda_3-\lambda_{4}\}$, where the $\lambda_i$'s are, in decreasing order, the square roots of the eigenvalues of the non-Hermitian matrix $\rho_{s,r}(t)\tilde{\rho}_{s,r}(t)$. The reduced density matrix  $\rho_{s,r}(t)$ is obtained after tracing over all the qubits in the channel besides sender and receiver, whereas $\tilde{\rho}_{s,r}(t)=(\sigma^y\otimes\sigma^y)\rho^*_{s,r}(t)(\sigma^y\otimes\sigma^y)$ and $\sigma^y$ is the Pauli spin-flip operator. After some algebraic manipulations, we reach the following expression for the concurrence between the two sites $s$ and $r$, i.e. between the two states $\bm{\ket s}$ and $\bm{\ket r}$\be
\mathcal{C}(t)=2\sin^2{\frac{\theta}{2}}\abs{f_{s,s}(t)}\abs{f_{s,r}(t)},\label{concu}\ee
where \be f_{s,n}(t)=\bm{\bra n}\bm{U}(t)\bm{\ket{s}}\label{prop}\ee is the probability amplitude of propagating the information from the sender site $s$ to the generic site $n$.
We note that  the concurrence Eq. (\ref{concu}) depends on the initial probability of having one excitation in the chain, from which the system can generate the entangled state. Indeed, in our case, this probability is given by $\sin^2(\theta/2)$. Therefore, we set $\theta=\pi$. Thus the concurrence $\mathcal{C}(t)$ becomes directly comparable to the transfer fidelity $F(t)=\abs{f_{s,r}(t)}^2$, when the latter is not averaged over the Bloch sphere  \footnote{ Usually the fidelity is
averaged over all the possible input states to give
$F(t)=\abs{f_{s,r}(t)}^2/6+\abs{f_{s,r}(t)}/3+1/2$.}.

We now examine the conditions under which the concurrence $\mathcal{C}(t)$ reaches the unitary value as a function of the transition amplitudes. For each $t$,  this search is subjected to the probability conservation constrain $
\abs{f_{s,s}(t)}^2+\abs{f_{s,r}(t)}^2+\Gamma_s^2(t)=1$,
where $\Gamma_s^2(t)=\sum_{i\neq(s,r)}\abs{f_{s,i}(t)}^2$ represents the dispersion of information in the channel, i.e. on all the sites besides $s$ and $r$. This condition leads to the individuation of
a curve of local maxima given by
$\abs{f_{s,s}(t)}^2=\abs{f_{s,r}(t)}^2$.
Hence, at the time $t$ we find a local maximum of $\mathcal{C}(t)$ if  \be \abs{f_{s,s}(t)}^2= \abs{f_{s,r}(t)}^2=\frac{1-\Gamma^2_s(t)}{2}, \label{locmax2}\ee
which becomes the  absolute maximum iff\be
\Gamma^2_s(t)=0\quad\mbox{and}\quad \abs{f_{s,s}(t)}^2= \abs{f_{s,r}(t)}^2=\frac{1}{2}.\label{absmax}\ee
Equations (\ref{locmax2}) and (\ref{absmax}) provide the conditions for generating maximally entangled states. In the next section we  show that these conditions are satisfied by systems which attain perfect state transfer by means of an effective 2-qubit dynamics.

\section{Entanglement generation and perfect state transfer}\label{ent}

It is convenient to recast the propagator in Eq. (\ref{prop}) in terms of the system eigenvalues $\{E_j\}$ and eigenvectors $\{\ket{\lambda_j}\}$ 
\be f_{s,n}(t)=\sum_{j=1}^N\bm{\bra{n}}\laj\rangle\langle\laj\bm{\ket{s}}\exp(-iE_jt)\ee
in order to recast Eq. (\ref{locmax2}) as
\be
 \left |\sum_{j=1}^N\abs{\sigma_j}^2e^{-iE_jt}\right |=
\left |\sum_{j=1}^N\sigma_j\rho_j^*e^{-iE_jt}\right|.
\label{propa}\ee The relevant quantities are \be
\sigma_j=\langle\laj\bm{\ket{s}},\,\,
\rho_j=\langle\laj\bm{\ket{r}},\,\,
\gaj=\sum_{i\neq (s,r)}\abs{\langle\laj\bm{\mid i\rangle}}^2,\label{gaj}\ee
which represent the projection of the $j$th eigenvector $\ket\laj$ on, respectively, the sender state, the receiver state, and the remaining system states.  For each $j$  the normalization condition $\sij+\roj+\gaj=1$ is fulfilled. 
If one looks for a solution of Eq. (\ref{locmax2}) independent of the size of the system $N$, i.e. scalable, and of related specific properties of the spectrum, then Eq. (\ref{propa}) provides a set of conditions for each $\abs{\sigma_j},\abs{\rho_j}$ namely \be \sij=\roj=\frac{1-\gaj}{2}\qquad\forall j,\label{sym}\ee
which represent exactly the local maximum conditions for perfect state transfer  \cite{Gualdi08} in case of a generic spectrum.

 Next, in order to satisfy also Eq. (\ref{absmax}), we consider
\be\Gamma^2_s(t)=\sum_{j=1}^N\abs{\sigma_j}^2\abs{\gamma_j}^2+
2\sum_{j<j'}\abs{\sigma_j}\abs{\sigma_{j'}}\abs{\gamma_{j,j'}}\cos\left[(E_j-E_{j'})t+\xi_{j,j'}\right], \label{gamma}\ee
where $\abs{\gamma_{j,j'}}=\sum_{i\neq s,r}\abs{\bra{\lambda_j}\bm{i}\rangle}\abs{\bra{\lambda_{j'}}\bm{i}\rangle}$ and $\xi_{j,j'}$ is a phase factor.  Obviously \be
\Gamma^2_s(t)\leq\sum_{j=1}^N\abs{\sigma_j}^2\abs{\gamma_j}^2+2\sum_{j<j'}\abs{\sigma_j}\abs{\sigma_{j'}}\abs{\gamma_{j,j'}},\label{cos}\ee
regardless to the specific spectrum of the system.
Moreover, due to the normalization $\abs{\gamma_{j,j'}}\leq\gaj$. Hence, we can further maximize Eq. (\ref{cos}), by writing \bea
\Gamma^2_{s}(t)&\leq&\sum_{j=1}^N\abs{\sigma_j}^2\abs{\gamma_j}^2+2\sum_{j<j'}\abs{\sigma_j}\abs{\sigma_{j'}}\gaj\nn\\
&\leq& N\sum_{j=1}^N\abs{\sigma_j}^2\abs{\gamma_j}^2\equiv N\Gamma_M.\label{gajj}\eea
Necessary and sufficient condition in order to have $\Gamma^2_s(t)=0$ is that $\abs{\gamma_j}^2=0$ for  each $j$ for which $\abs{\sigma_j}^2\neq 0$.  Now, we prove that  there can be only two eigenvectors for which $\abs{\sigma_j}^2\neq0$ and $\gaj=0$. Given the normalization constrain $\sum_j\gaj=N-2$, in order to minimize $\Gamma_M$ we need to  find its  lowest extreme, with respect to $\gaj$.
When we use the local maximum conditions, Eq. (\ref{sym}), we obtain
\be\Gamma_M=\sum_{j=1}^N\frac{1-\gaj}{2}\gaj.\label{gam}\ee The extremal point is then reached for $\gaj=(N-2)/N$  for which $\Gamma_M=(N-2)/N$.
 The absolute minimum, $\Gamma_M=0$, is attained  for  $N=2$, i.e. for a truly 2-qubit system. We note that for $N=2$ also $\gaj=0$.  Hence, to generate a maximally entangled state, when $N>2$, only two eigenvectors  must have finite projections on sender and receiver states and zero projections on the other system states.
 This means that the $N$-qubit Hilbert space of the system can be decomposed as $\mathcal{H}^{N\times N}=\mathcal{H}^{2\times2}_{(s,r)}\oplus\mathcal{H}^{N-2\times N-2}_{channel}$, which is precisely the perfect state transfer condition for a generic system \cite{Gualdi08}.

 \section{The isotropic Heisenberg model}\label{heis}

  We now study the properties and time behaviour of the concurrence $\mathcal{C}(t)$ for a finite linear chain of interacting spins according to the isotropic XYZ Heisenberg model\be
H=\sum_{i,j;i\neq j}^NJ_{i,j}(\bm S_i\cdot\bm S_j-3S^z_iS^z_j)\quad\mbox{with}\; J_{i,j}=\frac{C}{(a\abs{i-j})^{\nu}},\label{Hnu}\ee
where $\nu>0$, $a$ is the fixed inter-spin distance and $C$ is a model depending constant. In particular, the case $\nu=3$ corresponds to the dipolar coupling.
In the truly 2-qubit sender-receiver subspace the propagator of the excitation from sender to receiver is
\be\abs{f_{s,r}(t)}^2=\sin^2\left(\frac{\Delta}{2}t\right),\label{fid}\ee
where $\Delta\equiv(E_+ -E_-)=2J_{s,r}$, with $J_{s,r}$ being the coupling between  sender and receiver. Perfect state transfer takes place at the time $T=\pi/\Delta$  when $\abs{f_{s,r}(T)}^2=1$ [Fig. \ref{fig1}(c)].
The transmission distance affects only the transfer time  through the explicit expression of $J_{s,r}$.
This is also true for a $N$-qubit system whenever the dynamics is confined to the sender-receiver  subspace, regardless of the actual system size.
We note that  the maximum concurrence, Eq. (\ref{absmax}), is reached at the time $t=T/2$, when $\abs{f_{s,s}(t)}^2=\abs{f_{s,r}(t)}^2=1/2$ [Fig. \ref{fig1}(b)].
 We emphasize that entanglement generation between sender and receiver shares the same properties of the information transfer. Both concurrence and fidelity peak to the same maximum value regardless of  the distance between the two parties and of  the number of qubits in the channel.

The correspondence  between perfect state transfer and entanglement generation applies only to {\it non} mirror-periodic systems. Instead, for mirror-periodic systems \cite{Christandl04,Albanese04} perfect state transfer represents the mark of minimum concurrence between $s$ and  $r$. In these cases, the system evolves in a $N$-qubit space (see, for instance, the fictitious-particle representation in \cite{Christandl04,Albanese04}), such that $\gaj\neq0$ always, for each system eigenvector. This implies that $\Gamma_s^2(t)\neq0$, for each value of $t$ except for $T$, the time at which perfect state transfer takes place. At this time, however, $\abs{f_{s,s}(T)}^2=0$ and $\abs{f_{s,r}(T)}^2=1$, thus  the conditions expressed in Eq. (\ref{absmax}) are never simultaneously fulfilled. This fact is reasonable  from a physical point of view. In fact, in mirror-periodic systems,
perfect state transfer is reached at the time $T$ when all the $N$ eigenfrequencies of the system recombine coherently  and the message revives at the sender-mirroring point in the chain. This implies that all degrees of freedom of the system are strongly coupled. The formation of an entangled state, instead, takes place only when two degrees of freedom are strongly correlated and fully detached from the others.
Indeed $\gaj$, which represents the mixing between the sender-receiver subspace and the rest of chain,
 quantifies how closely a system resembles a 2-qubit one.

 \section{Numerical results}\label{num}

\begin{figure}\begin{center}
\includegraphics[width=9cm]{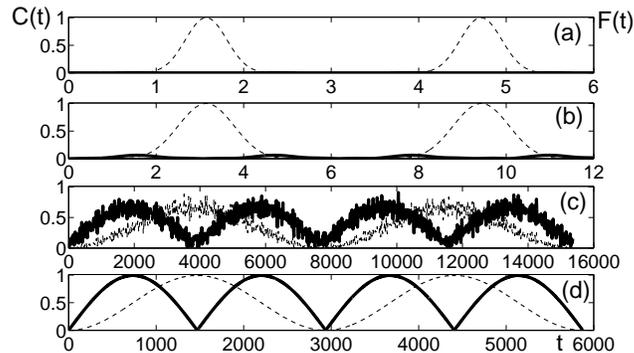}
\caption{\label{fig2}Concurrence (solid line) and fidelity (dashed line) as a function of time for mirror-periodic, (a) and (b), and generic, (c) and (d), spin chains. (a) Nearest neighbour interaction with $N=10$ sites. (b) Dipolar interaction with $N=6$ sites engineered according to \cite{Kay06}; (c) $N=10$ sites  and uniform filling; (d) $N=10$ sites  in the DH setup.
Time is measured in units of $a^3/C$  [see Eq. (\ref{Hnu})]. In case (a) the parameter $\lambda$ of Ref. \cite{Christandl04} is set equal to $2C/a^3$.
}\end{center}
\end{figure}

To provide a numerical evidence to our analytical results, in Fig. \ref{fig2} we plot the concurrence and the transfer fidelity for the interacting XYZ Heisenberg ferromagnetic qubit chain, Eq. (\ref{Hnu}). We compare the performances of mirror-periodic cases, (a) and (b), to those of systems with a generic spectrum, (c) and (d) . In Fig. \ref{fig2}(a), the mirror-periodic Hamiltonian for $N=10$ sites  is characterized by a nearest-neighbour interaction with a non-uniform $J$ coupling, chosen according to the procedure outlined in Refs. \cite{Christandl04,Albanese04}.
Plot (b) pertains to the example presented in Ref. \cite{Kay06} of a  mirror-periodic Hamiltonian with  dipole-like interaction and non-uniform coupling strength. We note that  in both cases the fidelity achieves the unitary value. The concurrence, instead, is practically zero in case (a), hence indistinguishable from the horizontal axis, and very close to zero in case (b). \\
Plot (c) depicts the behaviour of a uniform chain with dipolar interaction. We note that both fidelity and concurrence approximately achieve the same maximum value, though at different times. In this case the dynamics is approximately dominated by the two lowest-energy eigenvectors. Therefore $\gaj$ is small, but not neglibile. The truly 2-qubit dynamics is attained by
 the so-called double-hole (DH) model \cite{Gualdi08},  obtained from the same system by removing sender and receiver nearest-neighbouring spins [Fig. \ref{fig2}(d)].
 In this case $\gaj$ is  zero for any practical purpose. In other words, no information is left in the channel, but it periodically oscillates between sender and receiver. Therefore the DH chain not only attains perfect state transfer but also generates maximally entangled states. Also relevant for experimental implementations, it is the regular time-behaviour of both concurrence and fidelity, which greatly relaxes the time resolution requested for the measurement.

\begin{figure}\begin{center}\includegraphics[width=10cm]{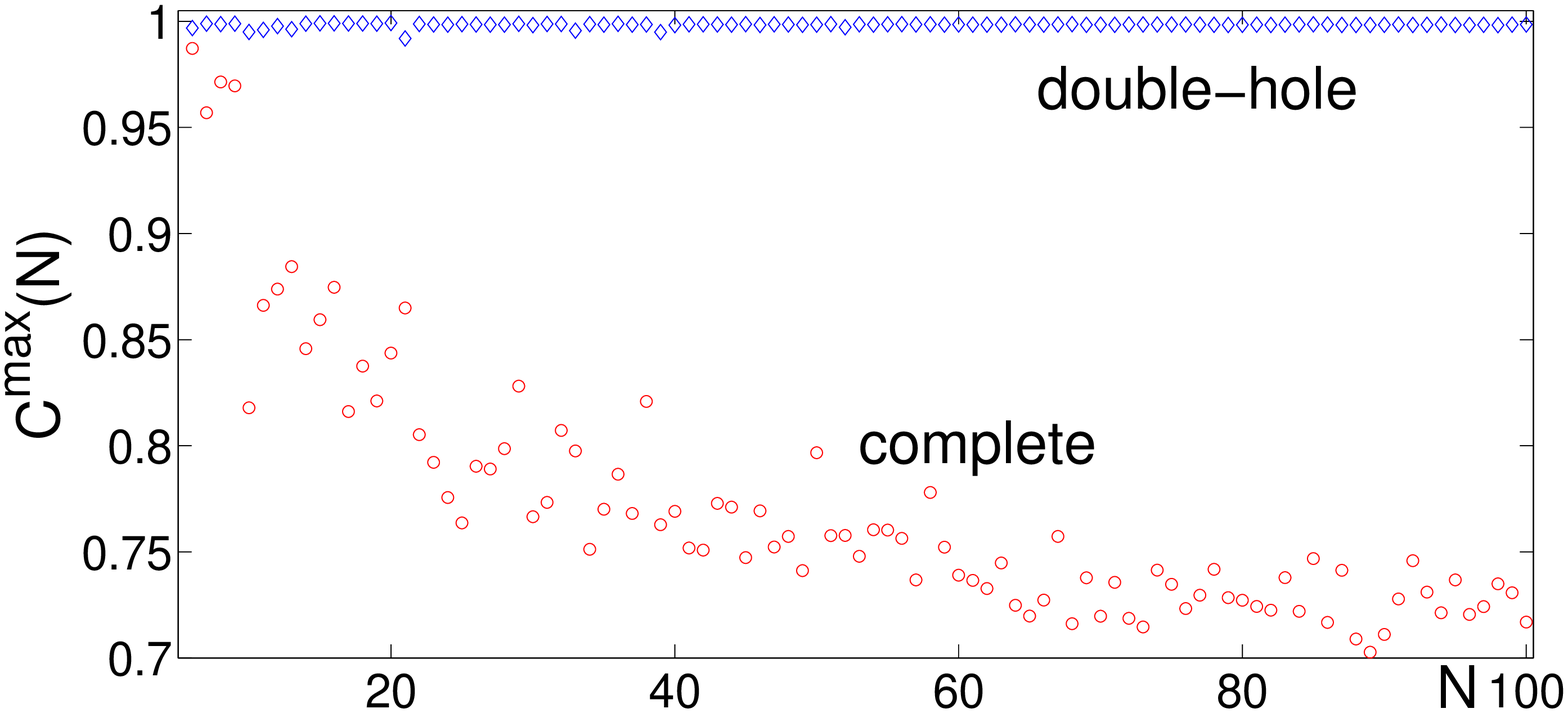}\caption{\label{fig3}(colour online) The maximum concurrence between sender and receiver as a function of the number $N$ of sites, for a complete chain (red circles) and for the DH chain (blue diamonds).}\end{center}\end{figure}

 In Fig. \ref{fig3} we plot the concurrence as a function of the distance between sender and receiver for a  chain of equally spaced qubit sites. In the case of the dipolar complete chain (uniform filling) the concurrence strongly decreases when increasing the number of spins between sender and receiver. This behaviour can be understood in terms of the spectral properties of the system. The longer the chain, the smaller the difference in energy between the two lowest eigenvectors and all the others. Therefore the  mixing between the sender-receiver subspace and the channel becomes less and less negligible    ($\gaj\neq0$). To restore the ideal 2-qubit dynamics it is sufficient to leave empty the sender and receiver nearest-neighbouring sites. Indeed, from Fig. \ref{fig3} we see that the concurrence of the DH system is practically insensitive to the distance between sender and receiver and to the number of spins in between.

\section{Conclusions}\label{concl}

 In this paper we have proved that  a ferromagnetic qubit system  can be used as a quantum channel not only for optimal state transfer but also for generating maximally entangled states over an arbitrary distance.  Starting from a long-range interacting spin chain, with a generic spectrum, we are able to confine the dynamics of the system to an effective 2-qubit subspace.
 Once the dynamics
is governed by these two qubits,
one can either retrieve the information with unitary fidelity or prepare the system in one of the two Bell
states $\bm{\ket{\Psi^{\pm}}}$. Our approach is fully scalable, since it only requires to leave empty the nearest-neighbouring sites of sender and receiver. Therefore, we can envisage a straightforward implementation of the proposed scheme with  strings of trapped particles \cite{Porras04,Ciaramicoli07, Ciaramicoli08, Johanning08, Haeffner08, Ospelkaus08,Marzoli08, Friedenauer08}.

\ack

 This research was supported by the European Commission through the Specific Targeted Research Project  QUELE and the  Integrated Project  FET/QIPC SCALA.

\end{document}